\documentclass[epj]{svjour}
\usepackage{graphics}
\newcommand{\be}{\begin{equation}}
\newcommand{\bea}{\begin{eqnarray}}
\newcommand{\ee}{\end{equation}}
\newcommand{\eea}{\end{eqnarray}}
\newcommand{\bpi}{\begin{picture}}
\newcommand{\bce}{\begin{center}}
\newcommand{\epi}{\end{picture}}
\newcommand{\ece}{\end{center}}
\newcommand{\D}{\displaystyle}
\def\chic#1{{\scriptscriptstyle #1}}

\def\g{\widetilde{{\rm I}\hspace{-0.07cm}\Gamma}}
\def\gt{\widetilde{\Gamma}^{\rm L}_{\nu\alpha\beta}}
\def\gnp{{\overline g}^2_{{\chic {\rm NP}}}}

\begin{document}
\title{On dynamical gluon mass generation}

\author{A.~C~.Aguilar\inst{1} \and J.~Papavassiliou\inst{2}
}                     

\institute{ Instituto de F\'{\i}sica Te\'orica,
Universidade Estadual Paulista,
Rua Pamplona 145,
01405-900, S\~ao Paulo, SP, Brazil. \and Departamento de F\'{\i}sica Te\'orica and IFIC, 
Universidad de Valencia-CSIC, 
E-46100, Burjassot, Valencia, Spain.}

\date{18 December 2006} 
\abstract{ The  effective gluon propagator constructed  with the pinch
technique  is  governed by  a  Schwinger-Dyson  equation with  special
structure  and  gauge  properties,   that  can  be  deduced  from  the
correspondence with the background field method.  Most importantly the
non-perturbative gluon self-energy is transverse order-by-order in the
dressed  loop   expansion,  and  separately  for   gluonic  and  ghost
contributions, a property which allows for a meanigfull truncation.  A
linearized  version  of  the  truncated  Schwinger-Dyson  equation  is
derived, using a vertex that  satisfies the required Ward identity and
contains massless poles.  The  resulting integral equation, subject to
a  properly  regularized constraint,  is  solved  numerically, 
and the main features of the solutions are briefly discussed.
} 

\maketitle
 It  is well-known  that one  of  the main  theoretical problems  when
dealing with Schwinger-Dyson (SD) equations is that they are built out
of  unphysical off-shell  Green's functions;  thus, the  extraction of
reliable physical information  depends crucially on delicate all-order
cancellations, which may be  inadvertently distorted in the process of
the truncation.   The truncation scheme  based on the  pinch technique
(PT)  ~\cite{Cornwall:1982zr,Cornwall:1989gv}   implements  a  drastic
modification at  the level  of the building  blocks of the  SD series.
The PT  enables the construction  of new, effective  Green's functions
endowed  with  very special  properties;  most  importantly, they  are
independent of  the gauge-fixing parameter, and  satisfy QED-like Ward
identities (WI)  instead of the usual  Slavnov-Taylor identities.  The
upshot of  this approach  would then be  to trade the  conventional SD
series for another, written in terms of the new Green's functions, and
then  truncate this  new series,  by  keeping only  a few  terms in  a
``dressed-loop''  expansion, maintaining  exact  gauge-invariance.  Of
central importance  in this context  is the connection between  the PT
and  the  Background  Field   Method  (BFM),  a  special  gauge-fixing
procedure  that preserves the  symmetry of  the action  under ordinary
gauge transformations with respect to the background (classical) gauge
field  $\widehat{A}^a_{\mu}$. As  a result,  the  background $n$-point
functions satisfy  QED-like all-order WIs.  The  connection between PT
and  BFM,   known  to  persist   to  all  orders
(last two articles in \cite{Cornwall:1989gv}),  
affirms   that  the
(gauge-independent) PT effective $n$-point functions coincide with the
(gauge-dependent) BFM $n$-point functions provided that the latter are
computed in the Feynman gauge. 
In this talk we report recent progress on the issue of gluon mass generation in the 
PT-BFM scheme~\cite{Aguilar:2006gr}.

We first define  
some basic quantities. There are two gluon propagators appearing 
in this problem, $\widehat{\Delta}_{\mu\nu}(q)$ and ${\Delta}_{\mu\nu}(q)$,
denoting the background and quantum gluon propagator, respectively. 
Defining ${\rm P}_{\mu\nu}(q)= \ g_{\mu\nu} - \frac{\D q_\mu
q_\nu}{\D q^2}$, we have that $\widehat{\Delta}_{\mu\nu}(q)$, 
in the  Feynman gauge is given by 
\begin{equation}
\widehat{\Delta}_{\mu\nu}(q)= {-\D i}\left[{\rm P}_{\mu\nu}(q)\widehat{\Delta}(q^2) + 
\frac{q_{\mu}q_{\nu}}{q^4}\right],
\label{prop_cov}
\end{equation}
The gluon self-energy, $\widehat{\Pi}_{\mu\nu}(q)$, has the form 
$\widehat{\Pi}_{\mu\nu}(q)={\rm P}_{\mu\nu}(q)\,\widehat{\Pi}(q^2)$, and 
$\widehat{\Delta}^{-1}(q^2) = q^2 + i\widehat{\Pi}(q^2)$. Exactly analogous definitions relate ${\Delta}_{\mu\nu}(q)$ with ${\Pi}_{\mu\nu}(q)$.

As is  widely known, in the conventional formalism  
the inclusion of ghosts is instrumental for  the   transversality   of
$\Pi^{ab}_{\mu\nu}(q)$, already at the level of the one-loop calculation.
On  the other  hand, in the  PT-BFM formalism,
due  to new Feynman  rules for  the vertices,  the one-loop  gluon and
ghost contribution are individually transverse \cite{Abbott:1980hw}.

\begin{figure}
\resizebox{0.5\textwidth}{!}{%
  \includegraphics{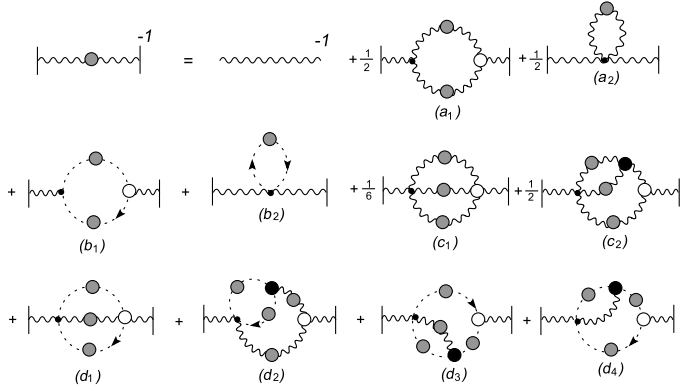}}
\caption{The SD equation for the gluon propagator in the BFM. 
All external legs (ending with a vertical line) are background gluons,
wavy lines with grey blobs denote
full-quantum gluon propagators, dashed lines with 
grey blobs are full-ghost propagators,  black dots are the  
BFM tree-level  vertices,  black blobs are the full conventional  vertices,   
and white blobs denote full three or four-gluon vertices with 
one external background leg.}
\label{fig:1}       
\end{figure}

As has been shown in~\cite{Aguilar:2006gr},
this crucial feature  persists  at the 
non-perturbative level, as a consequence of the simple WIs satisfied by
the full vertices appearing in 
the  diagrams  of Fig.(\ref{fig:1}), defining the BFM SD equation 
for $\widehat{\Delta}_{\mu\nu}(q)$~\cite{Sohn:1985em}.
Specifically, the  gluonic  and  ghost  sector  are
separately  transverse, within each individual order in the dressed-loop expansion.

Let us demonstrate this property for graphs $({\bf a_1})$ and $({\bf a_2})$, given by
\begin{eqnarray}
\widehat{\Pi}^{ab}_{\mu\nu}(q)
\big|_{{\bf a_1}} &=& 
\frac{1}{2} \, \int\!\! [dk]\,
\widetilde{\Gamma}_{\mu\alpha\beta}^{aex}
\Delta^{\alpha\rho}_{ee'}(k)
{\g}_{\nu\rho\sigma}^{be'x'}
\Delta^{\beta\sigma}_{xx'}(k+q)  \,,
\nonumber\\
\widehat{\Pi}^{ab}_{\mu\nu}(q)
\big|_{{\bf a_2}} &=&
\frac{1}{2} \,\int\!\! [dk]\,
\widetilde{\Gamma}_{\mu\nu\alpha\beta}^{abex}
\Delta^{\alpha\beta}_{ex} (k) \, ,
\label{groupa}
\end{eqnarray}
where $ [dk] =  d^d k/(2\pi)^d$ \, 
with $d=4-\epsilon$ the dimension of space-time.
By virtue of the BFM all-order WI 
\be
q_1^{\mu}{\g}_{\mu\alpha\beta}^{abc}(q_1,q_2,q_3) =
gf^{abc}
\left[\Delta^{-1}_{\alpha\beta}(q_2)
- \Delta^{-1}_{\alpha\beta}(q_3)\right] \,,
\label{3gl} 
\ee
and using the tree-level $\widetilde{\Gamma}_{\mu\nu\alpha\beta}$ given in~\cite{Abbott:1980hw},
we have 
\bea
q^{\nu} \widehat{\Pi}^{ab}_{\mu\nu}(q)
\big|_{{\bf a_1}} &=&
 C_A \, g^2 \delta^{ab} \,q_{\mu}  \, \int\!\! [dk]\,
\Delta^{\rho}_{\rho}(k) \,,
\nonumber\\
q^{\nu}\widehat{\Pi}^{ab}_{\mu\nu}(q)
\big|_{{\bf a_2}} &=&
- C_A \, g^2  \delta^{ab} \,q_{\mu}  \, \int\!\! [dk]\,
\Delta^{\rho}_{\rho}(k) \,,
\label{Transgroupa}
\eea
and thus, 
$q^{\nu} ( 
\widehat{\Pi}^{ab}_{\mu\nu}(q)\big|_{{\bf a_1}}+
\widehat{\Pi}^{ab}_{\mu\nu}(q)\big|_{{\bf a_2}}) = 0  \,.$
The importance of  this transversality property in the  context of SD
equation  is that  it  allows for  a  meaningful first  approximation:
instead of the  system of coupled equations involving  gluon and ghost
propagators,  one  may consider  only  the  subset containing  gluons,
without  compromising the  crucial property  of  transversality.
We will therefore study as the  first  non-trivial approximation  for
$\widehat{\Pi}_{\mu\nu}(q)$ the  diagrams $({\bf a_1})$ and $({\bf a_2})$.
Of course, we have no a-priori guarantee
that this particular subset is numerically dominant.
Actually, as has been argued in a series of SD studies, in the context
of the conventional Landau gauge it is the ghost sector that furnishes
the leading contribution~\cite{vonSmekal:1997is}  Clearly, it
is plausible  that this characteristic feature may  persist within the
PT-BFM scheme as  well, and we will explore this  crucial issue in the
near future.

The equation given in (\ref{groupa}) is not a genuine SD equation, in the sense
that it does not involve the unknown quantity $\widehat{\Delta}$ on both sides.
Substituting $\Delta \to \widehat{\Delta}$ on the RHS of (\ref{groupa})
(see discussion in \cite{Aguilar:2006gr}), we obtain
\bea
\widehat{\Pi}_{\mu\nu}(q) &=& 
\frac{1}{2}\, C_A \, g^2 \,
 \int\!  [dk]\,
\widetilde{\Gamma}_{\mu}^{\alpha\beta}
\widehat{\Delta}(k)�
{\g}_{\nu\alpha\beta} 
\widehat{\Delta}(k+q)  \nonumber \\
 &-&\, C_A \, g^2 \, \,d \, g_{\mu\nu}
\int\!  [dk] \, \widehat{\Delta}(k) \, ,�
\label{polar2}
\eea
with
$\widetilde\Gamma_{\mu\alpha\beta}= (2k+q)_{\mu}g_{\alpha\beta} -2q_\alpha g_{\mu\beta} + 
2 q_\beta g_{\mu\alpha}$,
and
\be
q^{\nu} {\g}_{\nu\alpha\beta} = 
\left[\widehat{\Delta}^{-1}(k+q) - \widehat{\Delta}^{-1}(k)\right] g_{\alpha\beta}\, .
\label{WID}
\ee 
We can then linearize the resulting SD equation,  by resorting to
the Lehmann representation for the scalar part of the gluon propagator~\cite{Cornwall:1982zr}
\begin{equation}
\widehat\Delta (q^2) = \int \!\! d \lambda^2 \, \frac{\rho\, (\lambda^2)}{q^2 - \lambda^2 + i\epsilon}\, ,
\label{lehmann}
\end{equation}
and setting on the first integral of the RHS of Eq. (\ref{polar2})
\begin{equation}
\widehat\Delta (k) {\g}_{\nu\alpha\beta}
\widehat\Delta (k+q) 
= \int \!\!   
 \frac{d \lambda^2 \, \rho\,(\lambda^2) \,\gt}{[k^2 - \lambda^2] [(k+q)^2 -  \lambda^2]}
\label{propvert}
\end{equation}
where ${\gt}$ must be such as to satisfy the tree-level WI
\be
q^{\nu} {\gt} =\left[(k+q)^2 - \lambda^2 \right] g_{\alpha\beta} - (k^2 -\lambda^2)g_{\alpha\beta}\,. 
\label{wigtt}
\ee
We propose the following form for the vertex
\bea
{\gt} &=&
\widetilde\Gamma_{\nu\alpha\beta}    +
    c_1 \left((2k+q)_{\nu} + \frac{q_{\nu}}{q^2}
\left[k^2 - (k+q)^2\right]\right)g_{\alpha\beta} 
\nonumber\\
 &+& \left( c_3 + \frac{c_2}{2\, q^2}\left[(k+q)^2 + k^2 \right]\right)
\left( q_\beta g_{\nu\alpha} - q_\alpha g_{\nu\beta} \right)
\label{vertpoles}
\eea 
which, due to the  presence of the massless poles,
allows the  possibility of infrared finite solution. 

Due to the QED-like WIs satisfied by the PT Green's functions, 
$\widehat\Delta^{-1}(q^2)$ absorbs all  
the RG-logs.
Consequently, the product 
${\widehat d}(q^2) = g^2 \widehat\Delta(q^2)$ forms a RG-invariant 
($\mu$-independent) quantity.
Notice however that Eq.(\ref{polar2}) does not encode the correct RG behavior: 
when written in terms of  
$\widehat d(q^2)$ it is not
manifestly $g^2$-independent, as it should.
In  order  to  restore  the  correct  RG  behavior
we use  the  simple
prescription proposed in~\cite{Cornwall:1982zr},
whereby we  substitute every $\widehat\Delta(z)$ appearing
on RHS of the SD by
\be
\widehat\Delta(z) \to \frac{g^2\, \widehat\Delta(z)}{\bar{g}^{2}(z)}
\equiv [1+ \tilde{b} g^2\ln(z/\mu^2)]\widehat\Delta(z) \,. 
\label{gratio1}
\ee
Then, setting $\tilde b \equiv \frac{10 \,C_A}{48\pi^2}$, $\sigma\,\equiv \, \frac{6\,(c_1+c_2)} {5}\,$,
$\gamma\,\equiv \,\frac{4+4\,c_1+3\,c_2} {5}\,$, we finally obtain
\bea
{\widehat d}^{\,-1}(q^2) &=& q^2 \Bigg\{ K^{\prime}
+ {\tilde b} 
\int^{q^2/4}_{0}\!\!\! dz\, \left(1-\frac{4z}{q^2}\right)^{1/2}
\frac{{\widehat d}(z)}{\overline{g}^2(z)}\Bigg\}
\nonumber\\
&+&\,
\gamma {\tilde b}
\int^{q^2/4}_{0}\!\!\!dz \,z \,
\left(1-\frac{4z}{q^2}\right)^{1/2}
\frac{{\widehat d}(z)}{\overline{g}^2(z)} 
\nonumber\\
&+& {\widehat d}^{\,-1}(0) \,,
\label{sd5}
\eea
\be
K^{\prime} = \frac{1}{g^2} - {\tilde b}  
\int^{\mu^2/4}_{0}\!\!\!dz \, 
\left(1+\gamma\,\frac{z}{\mu^2} \right)\, 
\left(1-\frac{4z}{\mu^2}\right)^{1/2}\,
\frac{{\widehat d}(z)}{\overline{g}^2(z)}\,,
\label{constreuc}
\ee
and
\be
{\widehat d}^{\,-1}(0) = 
 -  \frac{{\tilde b}\sigma}{\pi^2}
\int\! d^4 k \,\frac{{\widehat d} (k^2)}{\bar{g}^2(k^2)} \,.
\label{d01}
\ee 
It is easy to see now that Eq.(\ref{sd5}) yields the correct UV behavior, i.e.  ${\widehat d}^{\,-1}(q^2)= \tilde{b}\,q^2\ln(q^2/\Lambda^2)$.

When solving (\ref{sd5})
we will be interested in solutions that are qualitatively of the general form~\cite{Cornwall:1982zr} 
\be
{\widehat d}(q^2) = \frac{\gnp(q^2)}{q^2 + m^2(q^2)}\,,
\label{ddef}
\ee
where 
\be
\gnp(q^2) = \bigg[\tilde{b}\ln\left(\frac{q^2 + f(q^2, m^2(q^2))}{\Lambda^2}\right)\bigg]^{-1}\,,
\label{GNP}
\ee
$\gnp(q^2)$
represents a non-perturbative version of the 
RG-invariant effective charge of QCD:
in the deep UV it goes over to $\overline{g}^{2}(q^2)$, 
while in the deep IR it ``freezes''~\cite{Cornwall:1982zr,Aguilar:2002tc},
due to the presence of the function 
$f(q^2, m^2(q^2))$, whose form will be determined by fitting the numerical solution.  
The function $m^2(q^2)$ may be interpreted as a  
momentum dependent ``mass''. 
In order
to determine the  asymptotic behavior that Eq.(\ref{sd5}) predicts
for  $m^2(q^2)$ at large $q^2$,
we replace Eq.(\ref{ddef}) on both sides, set
$(1- 4z/q^2)^{1/2}\to 1$, obtaining self-consistency provided that 
\be
 m^2(q^2) \sim  m^2_{0} \ln^{-a} \left(q^2/\Lambda^2\right)\,, \quad\mbox{with}\quad a= 1+\gamma >0 \, . 
\label{uv_mass} 
\ee 

The  seagull-like contributions, 
defining ${\widehat d}^{\,-1}(0)$ in (\ref{d01}), 
are essential for obtaining IR   finite    solutions. 
However,  the   integral in (\ref{d01})  should  be  properly
regularized, in order to ensure the finiteness of such a mass term.  
Recalling that in dimensional regularization 
$\int\!\,[dk]/k^2 =0$, we rewrite the Eq.(\ref{d01}) (using (\ref{ddef})) as  
\bea
{\widehat d}^{\,-1}(0) &\equiv&   -  \frac{{\tilde b}\sigma}{\pi^2}
\int\! [dk] \bigg(\,\frac{\gnp(k^2)}{[k^2 + m^2(k^2)]\bar{g}^{2}(k^2)} -\frac{1}{k^2}\bigg)
\nonumber\\
&& \hspace{-1cm}=  \frac{{\tilde b}\sigma}{\pi^2} \int\! [dk] \frac{m^2(k^2)}{k^2\, [k^2 + m^2(k^2)]}\, \nonumber\\ 
&& \hspace{-1cm}+  \frac{{\tilde b^2}\sigma}{\pi^2} \int\! [dk]\, {\widehat d}(k^2)\,
\ln\left(1 + \frac{f(k^2, m^2(k^2))}{k^2}\right) \, .
\label{basreg}
\eea
The first integral converges provided that $m^2(k^2)$ falls asymptotically as $\ln^{-a} (k^2)$,  
with $a >1$, while
the second requires
that $f(k^2, m^2(k^2))$ should drop asymptotically at least as fast as $\ln^{-c} (k^2)$, with $c>0$. 
Notice that perturbatively $ {\widehat d}^{\,-1}(0)$ 
vanishes, because $m^2(k^2)=0$ to all orders, and, in that case, $f=0$ also.

\begin{figure}
\resizebox{0.5\textwidth}{!}{%
  \includegraphics{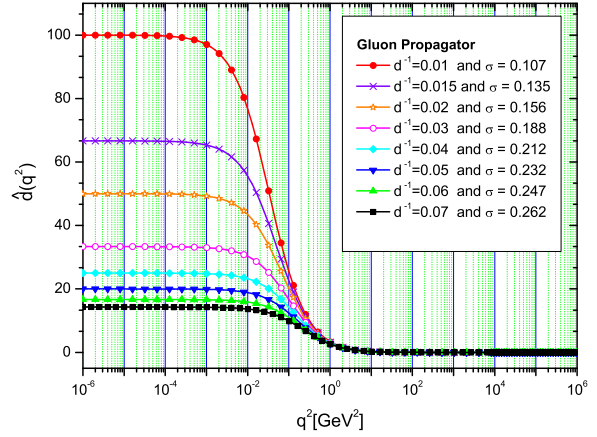}}
\caption{ Results for ${\widehat d}(q^2)$, for different values for ${\widehat d}^{\,-1}(0)$ 
(all in  $\;\mbox{GeV}^{\,2}$), and the corresponding values for $\sigma$.}
\label{fig:2}       
\end{figure}

\begin{figure}
\resizebox{0.5\textwidth}{!}{%
  \includegraphics{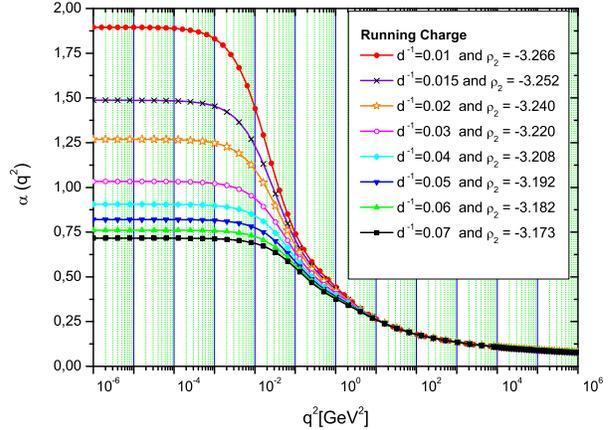}}
\caption{The running charge, $\alpha(q^2)$, corresponding 
to the gluon propagator of Fig.(\ref{fig:2}).}
\label{fig:3}       
\end{figure}

Solving numerically Eq.(\ref{sd5}), 
subject to the constraint of Eq.(\ref{d01}), we obtain solutions 
shown in Fig.(\ref{fig:2}); they can be fitted perfectly 
by means of a running coupling that freezes in the IR, shown in Fig.(\ref{fig:3}),
and a running mass that vanishes in the UV~\cite{Aguilar:2006gr}.
$\sigma$ is treated as a free parameter, whose values are 
fixed in such a way as to achieve compliance between Eqs.(\ref{sd5})-(\ref{d01}).

{\it Acknowledgments}:
This research was supported by Spanish MEC under the grant FPA 2005-01678 
and by  Funda\c{c}\~{a}o de Amparo  \`{a} Pesquisa do  Estado de
S\~{a}o  Paulo (FAPESP-Brazil)  through  the grant 05/04066-0.\,
J.P. thanks the organizers of QNP06 for their hospitality.

\end{document}